\documentclass[
reprint,
 amsmath,amssymb,
 aps,
]{revtex4-2}

\usepackage{graphicx}
\usepackage{dcolumn}
\usepackage{bm}
\usepackage{subcaption}
\usepackage{lipsum}
\usepackage{multirow}
\usepackage{siunitx}
\usepackage{nicefrac}
\usepackage{ulem}
\usepackage{xcolor}

\begin{document}

\title{Exciton radiative lifetimes in hexagonal diamond Ge and Si$_x$Ge$_{1-x}$ alloys}

\author{Michele Re Fiorentin}
\email{michele.refiorentin@polito.it}
\affiliation{%
Department of Applied Science and Technology, Politecnico di Torino, corso Duca degli Abruzzi 24, 10129 Torino, Italy
}%

\author{Michele Amato}
\affiliation{
 Université Paris-Saclay, CNRS, Laboratoire
de Physique des Solides, 91405 Orsay, France
}%

\author{Maurizia Palummo}%
\affiliation{%
INFN and Dipartimento di Fisica,
Università degli studi di Roma ``Tor Vergata", 00133 Rome,
Italy
}%

\date{\today}

\begin{abstract}
Recent reports of strong room-temperature photoluminescence in hexagonal diamond (2H) germanium stand in marked contrast to theoretical predictions of very weak band-edge optical transitions. Here we address radiative emission in 2H-Ge and related materials through a comprehensive investigation of their excitonic properties and radiative lifetimes, performing Bethe–Salpeter calculations on pristine and uniaxially strained 2H-Ge, 2H-Si$_x$Ge$_{1-x}$ alloys with $x=\nicefrac{1}{6},\,\nicefrac{1}{4},\,\nicefrac{1}{2}$, and wurtzite GaN as a reference.
Pristine 2H-Ge features sizable exciton binding energies ($\sim\!30\,\unit{meV}$) but extremely small dipole moments, yielding radiative lifetimes above $10^{-4}\,\unit{s}$. Alloying with Si reduces the lifetime by nearly two orders of magnitude, whereas a 2\% uniaxial strain along the $c$ axis induces a band crossover that strongly enhances the in-plane dipole moment of the lowest-energy exciton and drives the lifetime down to the nanosecond scale. Although strained 2H-Ge approaches the radiative efficiency of GaN, its much lower exciton energy prevents a full match.
These results provide the missing excitonic description of 2H-Ge and 2H-Si$_x$Ge$_{1-x}$, demonstrating that, even when excitonic effects are fully accounted for, the strong photoluminescence reported experimentally cannot originate from the ideal crystal. 
\end{abstract}

\maketitle


\section{Introduction}
\label{sec:intro}
The hexagonal diamond phase (lonsdaleite, 2H in the Ramsdell notation~\cite{Ramsdell_AmerMiner_1947}) of germanium has recently been grown in a variety of structures, from nanowires~\cite{Hauge_NanoLett_2017,Vincent_NanoLett_2018,Fadaly_Nature_2020} and nanobranches~\cite{Li_Nanotech_2023,Tizei_NanoLett_2025,Lamon_NanoLett_2025}, to quantum wells~\cite{Peeters_NatComm_2024,Peeters_PRB_2025}. The successful stabilization of this metastable phase in nanostructures has established 2H-Ge as a promising group-IV material for optoelectronic applications. The strong room-temperature photoluminescence (PL) observed by Fadaly et al.~\cite{Fadaly_Nature_2020} in 2H-Ge and 2H-Si$_x$Ge$_{1-x}$ nanowires further suggested the possibility of an intrinsic, efficient light emitter within this polytype. This result generated considerable interest by revealing an unexpected route to strong optical emission in group-IV materials, a class traditionally limited by the indirect band gap of the cubic diamond phase. Achieving optical emitters, lasers, detectors, and integrated photonic components remains a long-standing goal in the development of silicon photonics~\cite{Soref_IEEE_2006,Soref_NatPhotonics_2010,Priolo_NatNanotech_2014,Thomson_JOptics_2016,vanTilburg_JApplPhys_2023}. In this context, the apparent emergence of a direct-gap light emitter based solely on group-IV elements has renewed both scientific and technological interest in 2H-Ge.

Despite this excitement, theoretical analyses revealed inconsistencies between the reported strong PL and the predicted optical properties of 2H-Ge ~\cite{vanLange_ACSPhoton_2024,Broderick_arXiv_2024,Tizei_NanoLett_2025}. First-principles calculations consistently show that the fundamental $\Gamma$-point transition in 2H-Ge carries extremely weak oscillator strength. R\"{o}dl et al.~\cite{Rodl_PRM_2019} demonstrated that the dipole matrix element between the valence-band maximum (VBM) and the conduction-band minimum (CBm) at $\Gamma$ is strongly suppressed, classifying 2H-Ge as a pseudo-direct semiconductor. Optical absorption and electron energy-loss spectroscopy (EELS) further confirmed the absence of pronounced features at the electronic band edge~\cite{Rodl_PRM_2019,Tizei_NanoLett_2025}, revealing a nearly “dark” transition. Additional first-principles works~\cite{Rodl_PRM_2021,Borlido_PRM_2023} have reached similar conclusions, suggesting that the reported PL is not intrinsic and may instead arise from extrinsic mechanisms such as defects, morphology, or local strain fields \cite{Fadaly_Nature_2020,vanLange_ACSPhoton_2024,Belabbes_PSS_2022,Rovaris_ACSApplNanoMater_2024}.

Existing optical calculations on 2H-Ge have primarily focused on the independent-particle (IP) regime~\cite{Rodl_PRM_2019,Borlido_PRM_2023}, evaluating dipole transitions and PL assuming free electron–hole recombination. While such an approach captures essential characteristics, including the vanishing dipole matrix elements at the $\Gamma$ point, it neglects exciton formation and their impact on radiative recombination. Previous analyses have generally assumed a small exciton binding energy, by analogy with cubic Ge (3C-Ge), where $E_b\!\sim\!4\,\unit{meV}$~\cite{Madelung_Handbook}, implying that excitonic effects would be negligible at finite temperature and that an IP description would suffice. Hence, despite the growing interest in polytypic group-IV materials, a comprehensive study of the excitonic properties of 2H-Ge, SiGe alloys, and strained hexagonal variants has so far been lacking.

In this work, we aim to fill this gap by performing \textit{ab initio} calculations based on Density Functional Theory (DFT) and the Bethe-Salpeter equation (BSE) on pristine 2H-Ge, on 2H-Ge under uniaxial strain along the $c$ axis, a strategy previously shown to modify band ordering and enhance light emission~\cite{Rodl_PRM_2021}, and on representative 2H-Si$_{x}$Ge$_{1-x}$ alloys, with $x=\nicefrac{1}{6},\,\nicefrac{1}{4},\,\nicefrac{1}{2}$. We compute absorption spectra, exciton binding energies, dipole moments, and intrinsic radiative lifetimes, thereby providing a comprehensive excitonic description of these systems. This provides the first excitonic investigation of 2H-Ge--based materials, going beyond the IP analyses available so far and supplying an intrinsic optical benchmark long missing in this field.

We find sizable exciton binding energies ($\sim\!30\,\unit{meV}$ in pristine 2H-Ge), confirming that the excitonic picture is the most appropriate to describe the optical properties of these systems. 
Our results demonstrate that pristine 2H-Ge has intrinsically very long radiative lifetimes, exceeding $10^{-4}\,\unit{s}$ at low temperature, highlighting its pseudo-direct character. Alloying with Si moderately enhances the dipole strength by lifting symmetry-imposed constraints, whereas uniaxial strain produces a far more dramatic enhancement, reducing the radiative lifetime by more than five orders of magnitude and approaching the nanosecond regime. The strained system displays dipole moments and radiative efficiencies that approach those of wurtzite GaN, our reference efficient wide-bandgap emitter.

By combining exciton binding energies, dipole moments, oscillator strengths, and radiative lifetimes, our analysis provides a complementary theoretical perspective that reinforces previous findings of intrinsically weak optical activity in 2H-Ge. Our results substantiate the conclusion that the strong PL observed experimentally is likely driven by extrinsic mechanisms, while identifying strain engineering as a viable and highly effective route to enhance intrinsic light emission in 2H-Ge and reconcile theory and experimental results. The strain levels required to induce such effects are not unrealistic in nanowire geometries, where large elastic deformations can be sustained, with tensile strain approaching $\sim\!1.5\%$ reported in Ge nanowires and even larger values achievable, more generally, in semiconductor nanowires \cite{Guilloy_NanoLett_2015,Zhang_SciAdv_2016}.

The article is organized as follows. In Section~\ref{sec:methods} we describe the computational methods employed for the calculation of the electronic and excitonic properties, in Section~\ref{sec:results} we present our results on the materials and their exciton radiative lifetimes. Finally, in Section~\ref{sec:conclusion} we summarize our findings and draw our conclusions.

\section{Methods}
\label{sec:methods}
Ground-state calculations are carried out within plane-wave DFT, as implemented in the Quantum ESPRESSO code \cite{QE-2009,QE-2017,QE-2020}, with norm-conserving pseudopotentials, the PBEsol functional \cite{Perdew_PRL_2008}, and an 80 Ry kinetic-energy cutoff. Equilibrium structures are obtained through variable-cell optimization with increased cutoff value. 

To assess the behavior of hexagonal diamond SiGe alloys, we consider representative configurations built in a hexagonal rotated $\sqrt{3}\times\sqrt{3}\times2$ supercell containing 24 atoms. This size allows us to introduce compositional disorder without resorting to substantially more expensive larger supercells. We focus on Si$_{x}$Ge$_{1-x}$ alloys with $x=\nicefrac{1}{6},\,\nicefrac{1}{4},\,\nicefrac{1}{2}$, where atomic randomness is generated through the special quasi-random structure (SQS) method \cite{Zunger_PRL_1990,van_de_Walle_Calphad_2013}, which reproduces, already for relatively small cells, the radial correlation functions of an ideal random alloy. This model does not capture the full statistics of the alloys, which would require ensemble approaches or larger cluster sets such as those used in Ref. \cite{Borlido_PRM_2023}. Instead, it serves as a controlled, representative example to illustrate the effects on the radiative lifetime of mixing Si and Ge in the hexagonal phase. More details on the optimized structures are reported in the Supplemental Material, Section~S1.

To reduce the computational cost while obtaining reliable electronic band structures of 2H-Ge and related systems, we employ a $J$-parameter correction within the Liechtenstein formalism of DFT+$U$+$J$ \cite{Liechtenstein_PRB_1995}. In this scheme, the $J$ term introduces an on-site exchange interaction on the Ge $4p$ orbital subspace, which predominantly forms the valence-band edge. This adjustment lowers the energy of the $p$-derived valence states relative to the largely $s$-like conduction states, thereby opening the bandgap. As a result, a finite positive gap between the valence and conduction bands of 2H-Ge is recovered, whereas standard DFT incorrectly predicts a negative bandgap. The bandgap can be tuned to match experimental and theoretical reference values by varying the $J$ parameter; further details on its optimization are provided in the Supplemental Material, Fig.~S2(a).

We stress that the $J$ values used here are effective parameters within the projector-defined DFT+$J$ scheme, calibrated to reproduce the HSE06 bandgaps and band-edge dispersions. In this work, they are employed as a computational device to reduce the overall computational cost, without assigning them a direct interpretation in terms of physical Hund’s exchange parameters~\cite{Georges_AnnRev_2013}.

\begin{figure}[h]
    \centering
    \includegraphics[width=0.7\linewidth]{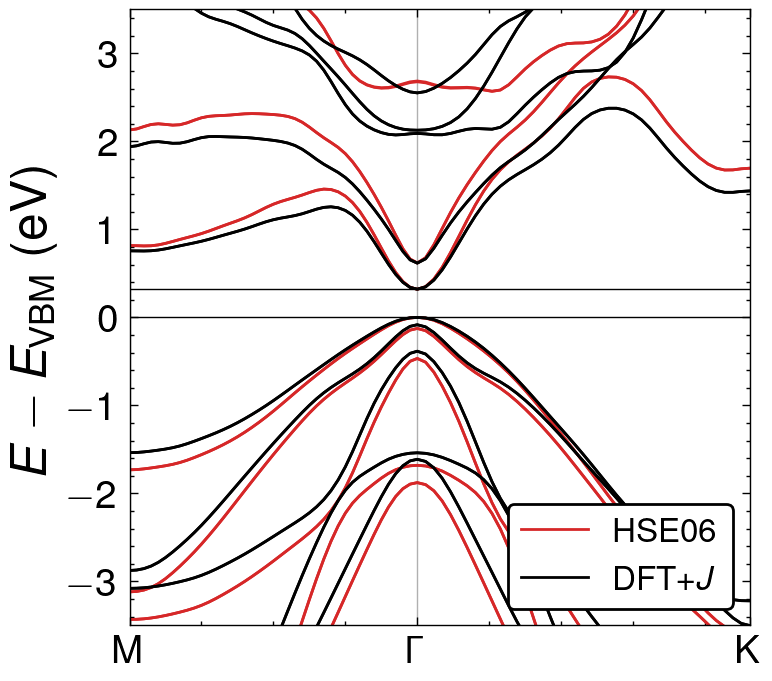}
    \caption{Electronic band structure of 2H-Ge near the band extrema, computed with the HSE06 hybrid functional (black lines) and within the DFT+$J$ method, with $J=23$~eV (red lines).}
    \label{fig:bands_HSE_DFTJ}
\end{figure}
In Fig.~\ref{fig:bands_HSE_DFTJ} we show the band structure of 2H-Ge computed with the Heyd-Scuseria-Ernzerhof range-separated hybrid functional (HSE06) \cite{Heyd_JChemPhys_2003} (black lines) and with DFT+$J$, using $J=23$~eV applied to the Ge $4p$ orbitals and no Hubbard $U$ correction (red lines).

The chosen value of $J$ reproduces the HSE06 bandgap $E_g = 0.32$~eV, consistent with previously reported values~\cite{De_JPhysCondMat_2014,Amato_JPCC_2017,Rodl_PRM_2019,Tizei_NanoLett_2025}, and yields a band structure in close agreement with the hybrid-functional result, in particular near the band extrema. This approach delivers HSE06-level accuracy at a fraction of the computational cost, enabling dense Brillouin-zone (BZ) sampling that would be impractical with hybrid functionals.

Throughout this work, the electronic band structures of 2H-Ge are computed using the $J = 23$~eV correction applied to the Ge $4p$ orbitals. For the 2H–Si$_{\nicefrac{1}{6}}$Ge$_{\nicefrac{5}{6}}$ and 2H–Si$_{\nicefrac{1}{4}}$Ge$_{\nicefrac{3}{4}}$ alloys, a larger value of $J = 40$ eV is required to match the corresponding HSE06 gap at that composition, reflecting the different local chemical environment in the alloy. For the Si-rich alloy 2H-Si$_{\nicefrac{1}{2}}$Ge$_{\nicefrac{1}{2}}$, a $J=40$~eV correction is also applied to the Si $3p$ orbitals to match the HSE06 bandgap. See the Supplemental Material, Fig.~S2(b) and (c), for more details. All calculations include spin–orbit coupling (SOC).

The electronic structures serve as the starting point for the optical-property calculations, obtained by solving the BSE with the YAMBO code \cite{Marini_ComputPhysComm_2009,Sangalli_JPhysCondMat_2019}, and the optimized diagonalization routines available in its Lumen branch~\cite{Lumen}. For 2H-Ge, we sample the BZ using a $21\times21\times14$ $k$-point grid, yielding 606 $k$-points in the irreducible BZ (6174 in the full BZ). We include 80 bands in the static screening, and apply a 2~Ry cutoff on the screened interaction. Additional details on the convergence of the $k$-point mesh, the number of bands included in the dielectric screening and the size of the static screening matrix are reported in the Supplemental Material, Fig.~S3(a) and (b). The BSE kernel is built using 6 valence and 4 conduction bands, accounting for SOC. For the GaN benchmark, we start from the plain DFT band structure corrected with a scissor shift to reproduce the HSE06 bandgap, and employ analogous parameters for the BSE kernel.

The diagonalization of the Bethe-Salpeter Hamiltonian yields the exciton dipole moment of state $S$,
\begin{equation}
\label{eq:dipoles}
\mu_{S,\alpha}=\sum_{cv\mathbf{k}}w_\mathbf{k}A^{S}_{cv\mathbf{k}}\langle c\mathbf{k}|r_\alpha|v\mathbf{k}\rangle,
\end{equation}
in the length gauge, where $A^S_{cv\mathbf{k}}$ is the exciton eigenvector of state $S$, and $\alpha$ denotes the in-plane ($\bot c$) or out-of-plane ($\| c$) components. 

The dipole moment also provides the dimensionless oscillator strength of exciton $S$,
\begin{equation}
    \label{eq:oscillator}
    f_{S,\alpha}=\frac{2m_0E_S}{\hbar^2e^2}|\mu_{S,\alpha}|^2,
\end{equation}
where $m_0$ is the electron rest mass and $E_S$ is the energy of exciton $S$.
Exciton dipole moments enter the radiative lifetime of state $S$ as given by \cite{Chen_PRB_2019,Jhalani_JPhysCondMat_2020},
\begin{align}
\label{eq:tauS}
    \tau_S(T)=&\left(\frac{2M_{\bot c}^{2/3}M_{\|c}^{1/3}c^2k_BT}{E_S^2}\right)^{3/2}\\
&\times\frac{\varepsilon_0\hbar V}{\sqrt{\pi\varepsilon_{\bot c}} e^2\left[\left(\frac{2\varepsilon_{\|c}}{3\varepsilon_{\bot c}}+2\right)|\mu_{S,\bot c}|^2+\frac{8}{3}|\mu_{S,\|c}|^2\right]},\nonumber
\end{align}
where $V$ is the simulation-cell volume, and $M_{\bot c},\,\varepsilon_{\bot c}$ and $M_{\| c},\,\varepsilon_{\| c}$ are the exciton mass and the optical dielectric constant for the in-plane and out-of-plane directions, respectively. 
The exciton effective masses entering Eq~(\ref{eq:tauS}) are approximated as the sum of the electron and hole effective masses along the corresponding crystallographic directions. The carrier effective masses are extracted from the DFT+$J$ band structure by fitting the band dispersion near the extrema using the \texttt{effmass} Python package \cite{Whalley2018}.
The overall radiative lifetime of the material at temperature $T$ is further obtained through a thermal average on the excitonic states \cite{Palummo_NanoLett_2015},
\begin{equation}
\label{eq:tau_avg}
    \langle\tau\rangle(T)=\left(\frac{\sum_S 1/\tau_S(T) e^{-E_S/k_BT}}{\sum_S e^{-E_S/k_BT}}\right)^{-1}.
\end{equation}
Eq.~(\ref{eq:tauS}) assumes a Maxwell–Boltzmann distribution of the exciton center-of-mass momenta (CMM), i.e. that CMM states are rapidly redistributed by scattering (phonons, disorder) so that the light cone remains thermally populated. Under this approximation $\tau_S(T)\propto T^{3/2}$, which formally vanishes as $T\rightarrow 0$. Yet, this low-temperature limit is not physical: once scattering into the light cone freezes out (below the light-cone energy scale, $\lesssim 1$~K), the model ceases to apply and the actual lifetime increases rather than going to zero, since excitons with zero CMM cannot radiatively recombine in 3D solids due to momentum conservation \cite{Chen_PRB_2019}. All temperatures considered here lie above this low-temperature crossover. We also note that radiative lifetimes computed within the IP approximation \cite{Delerue_PRB_1993,Rodl_PRM_2019,Fadaly_Nature_2020,Broderick_arXiv_2024} do not include this light-cone constraint and therefore remain finite as $T\rightarrow0$. The excitonic lifetimes evaluated here describe a different regime where momentum-conserving recombination is governed by the CMM distribution and exhibits the characteristic $T^{3/2}$ dependence.

\section{Results and Discussion}
\label{sec:results}
\subsection{Electronic structure}
\begin{figure*}[t]
\centering
\begin{subfigure}[b]{0.33\linewidth}
    \includegraphics[width=\linewidth]{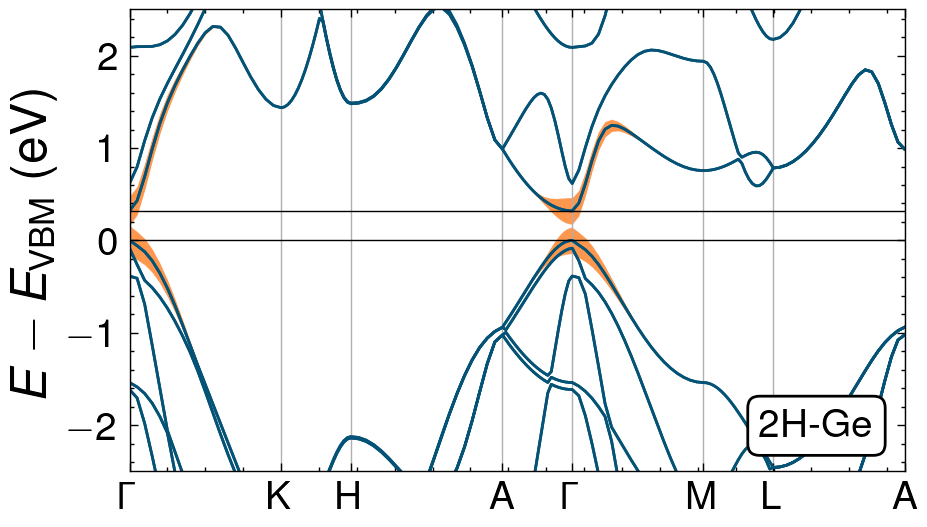}
    \caption{}
\end{subfigure}
\hspace{.2cm}
\begin{subfigure}[b]{0.31\linewidth}
    \includegraphics[width=\linewidth]{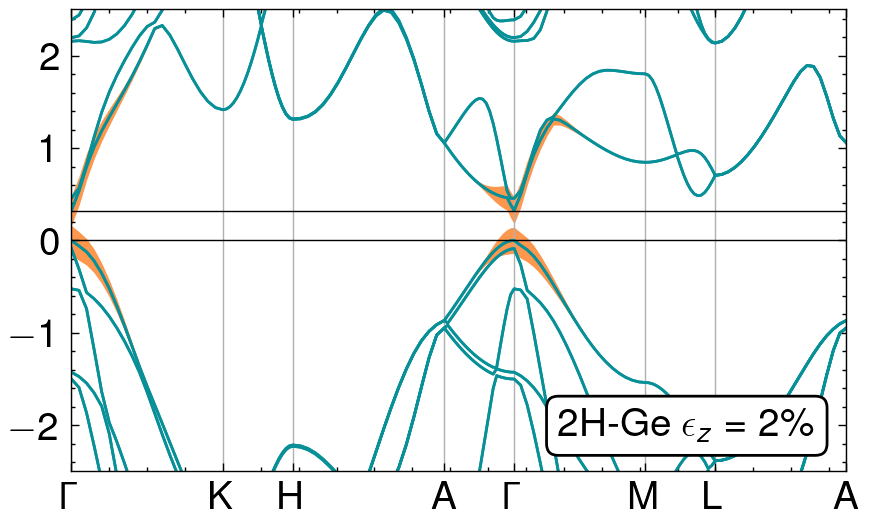}
    \caption{}
\end{subfigure}
\hspace{.2cm}
\begin{subfigure}[b]{0.31\linewidth}
    \includegraphics[width=\linewidth]{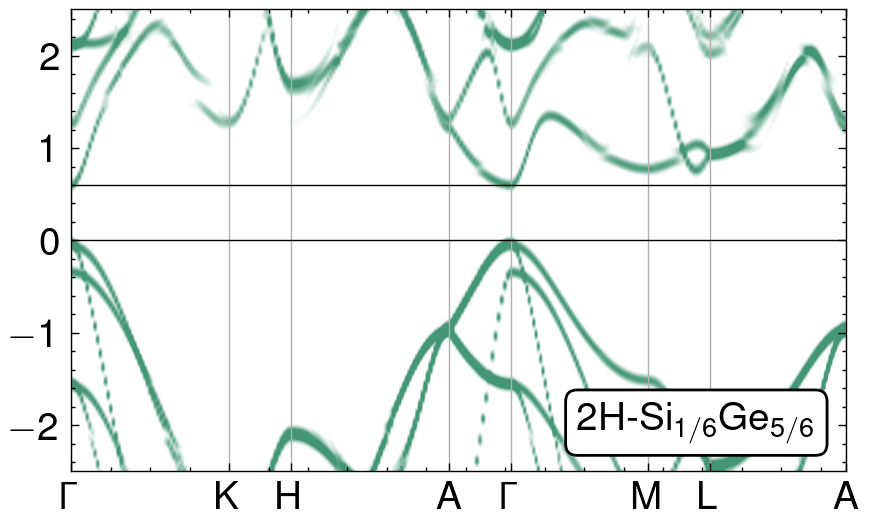}
    \caption{}
\end{subfigure}\\
\begin{subfigure}[b]{0.33\linewidth}
    \includegraphics[width=\linewidth]{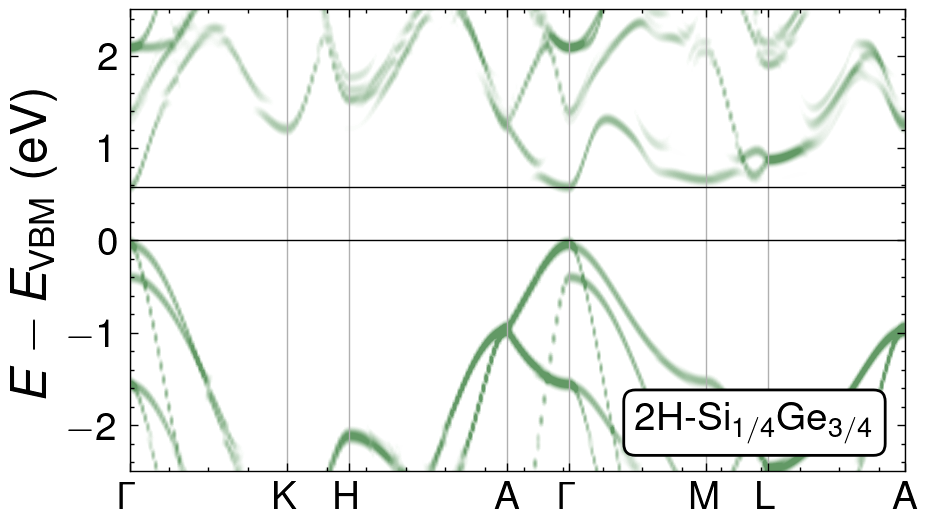}
    \caption{}
\end{subfigure}
\hspace{.2cm}
\begin{subfigure}[b]{0.31\linewidth}
    \includegraphics[width=\linewidth]{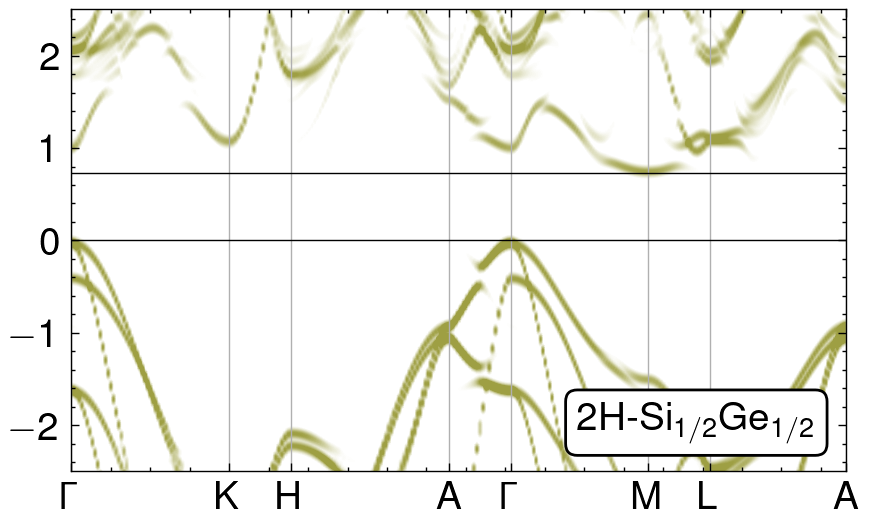}
    \caption{}
\end{subfigure}
\hspace{.2cm}
\begin{subfigure}[b]{0.31\linewidth}
    \includegraphics[width=\linewidth]{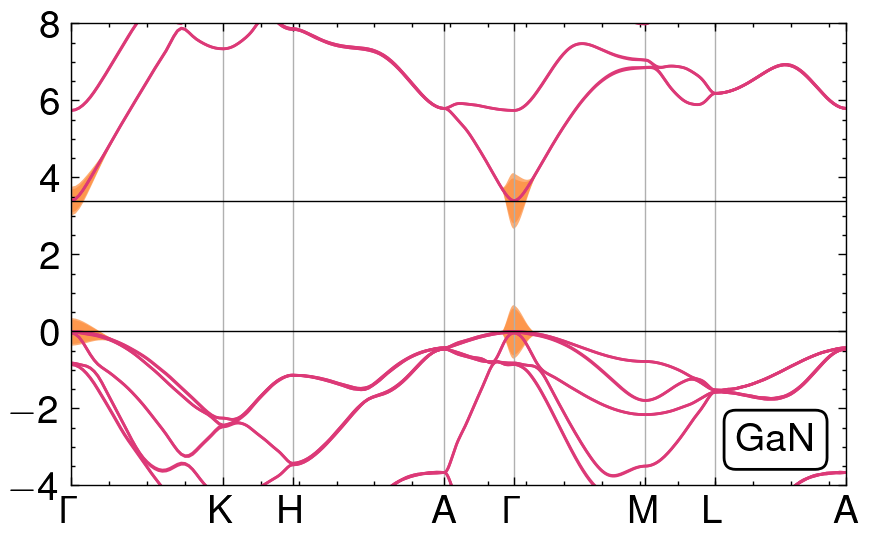}
    \caption{}
\end{subfigure}
\caption{Band structures of the studied materials: (a) pristine 2H-Ge, (b) 2H-Ge under $\epsilon_z=2\%$ uniaxial strain along the $c$ axis, (c) 2H-Si$_{\nicefrac{1}{6}}$Ge$_{\nicefrac{5}{6}}$ alloy, unfolded onto the 2H-Ge BZ, (d) 2H-Si$_{\nicefrac{1}{4}}$Ge$_{\nicefrac{3}{4}}$ alloy, unfolded, (e) 2H-Si$_{\nicefrac{1}{2}}$Ge$_{\nicefrac{1}{2}}$ alloy, unfolded, (f) GaN. The orange shading highlights the electronic states contributing to the lowest-energy exciton state. Horizontal lines mark the electronic bandgaps.}
\label{fig:bands}
\end{figure*}

The electronic band structure of 2H-Ge obtained with the DFT+$J$ method is shown in Fig.~\ref{fig:bands}(a). The orange shading highlights the electronic states that contribute most to the lowest excitonic state. The lowest-energy transitions occur near the $\Gamma$ point, from the VBM to the CBm. These transitions dominate the radiative properties of the material and are known to be particularly weak \cite{Rodl_PRM_2021}. The bandgap and the effective masses of electrons at the CBm and of holes at the VBM, both along $c$ ($\Gamma\!\rightarrow\!A$) and perpendicular to $c$ ($\Gamma\!\rightarrow\!M$), are reported in Tab.~\ref{tab:bands}, in good agreement with previous results~\cite{Rodl_PRM_2019}.

\begin{table}[h]
\caption{%
Electronic structure parameters. Bandgaps $E_g$ are in eV, and effective masses are expressed in units of the electron rest mass. For 2H-Si$_{\nicefrac{1}{2}}$Ge$_{\nicefrac{1}{2}}$, the direct gap at the $\Gamma$ point and the electron effective masses at the $M$ point are reported in brackets. 
}
\label{tab:bands}
\begin{ruledtabular}
\begin{tabular}{lccccc}
& $E_g$ & $m^{*}_{h,\bot c}$ & $m^{*}_{h,\| c}$ & $m^{*}_{e,\bot c}$ & $m^*_{e,\| c}$ \\
\colrule
2H-Ge & 0.32 & 0.147 & 0.488 & 0.090 & 1.160\\
2H-Ge $\epsilon_z=2\%$ & 0.32 & 0.119 & 0.499 & 0.038 & 0.031\\
2H-Si$_{\nicefrac{1}{6}}$Ge$_{\nicefrac{5}{6}}$ & 0.60 & 0.098 & 0.531 & 0.097 & 1.088\\
2H-Si$_{\nicefrac{1}{4}}$Ge$_{\nicefrac{3}{4}}$ & 0.58 & 0.121 & 0.698 & 0.097 & 1.774\\
\multirow{2}{8em}{2H-Si$_{\nicefrac{1}{2}}$Ge$_{\nicefrac{1}{2}}$} & 0.67 & \multirow{2}{2em}{0.226} & \multirow{2}{2em}{0.505} & 0.137 & 0.698\\
& (1.00) & & & (0.122) & (0.146)\\[5pt]
GaN & 3.40 & 0.164 & 2.418 & 0.144 & 0.161
\end{tabular}
\end{ruledtabular}
\end{table}

Fig.~\ref{fig:bands}(b) shows the band structure of 2H-Ge under 2\% uniaxial strain along the $c$ axis. As discussed in Ref.~\cite{Rodl_PRM_2021}, strain induces a crossover between the $\Gamma^-_{7c}$ state (CBm+1 in pristine 2H-Ge) and the $\Gamma^-_{8c}$ state (CBm in the unstrained system), which become, respectively, CBm and CBm+1 in the strained material. 
The value of 2\% strain should be regarded as a representative case where the band crossover is clearly established. Previous theoretical work indicates that the crossover may occur already at slightly lower strain ($\sim\!1–1.5\%$), depending on the computational approach \cite{Rodl_PRM_2021,Mayengbam_RSCAdv_2023}.
Although the bandgap changes only marginally, this crossover strongly affects the optical response: the lowest-energy exciton becomes dominated by much stronger optical transitions. The electron effective masses at the band edges (Tab.~\ref{tab:bands}) are significantly reduced compared to the unstrained case.

The band structures of the 2H-Si$_{x}$Ge$_{1-x}$ alloys, unfolded onto the 2H-Ge BZ, are shown in Figs.~\ref{fig:bands}(c)--(e). In agreement with previous calculations~\cite{Wang_APL_2021}, for $x=\nicefrac{1}{6}$ and $\nicefrac{1}{4}$, the alloys retain a direct gap, which increases to about 0.6~eV, while for $x=\nicefrac{1}{2}$ the bandgap is indirect between the $\Gamma$ and the $M$ points, see also Tab.~\ref{tab:bands}. The overall dispersions remain quite close to that of pristine 2H-Ge, except for the CBm+1 state, which is pushed upward in energy with increasing Si content, to roughly 1.2~eV, 1.5~eV and 1.8~eV above the VBM for $x=\nicefrac{1}{6},\,\nicefrac{1}{4}$ and $\nicefrac{1}{2}$, rispectively.


Finally, Fig.~\ref{fig:bands}(f) displays the band structure of GaN, used here as a benchmark. GaN features a much larger bandgap, with both VBM and CBm located at $\Gamma$, and the lowest-energy exciton dominated by the zone-center transitions highlighted in the figure. The computed effective masses, reported in Tab.~\ref{tab:bands}, are in good agreement with previous results in the literature \cite{Suzuki_PRB_1995}.

\subsection{Exciton dipoles and oscillator strengths}
\begin{figure*}[t]
\centering
\begin{subfigure}[b]{0.49\linewidth}
    \includegraphics[width=\linewidth]{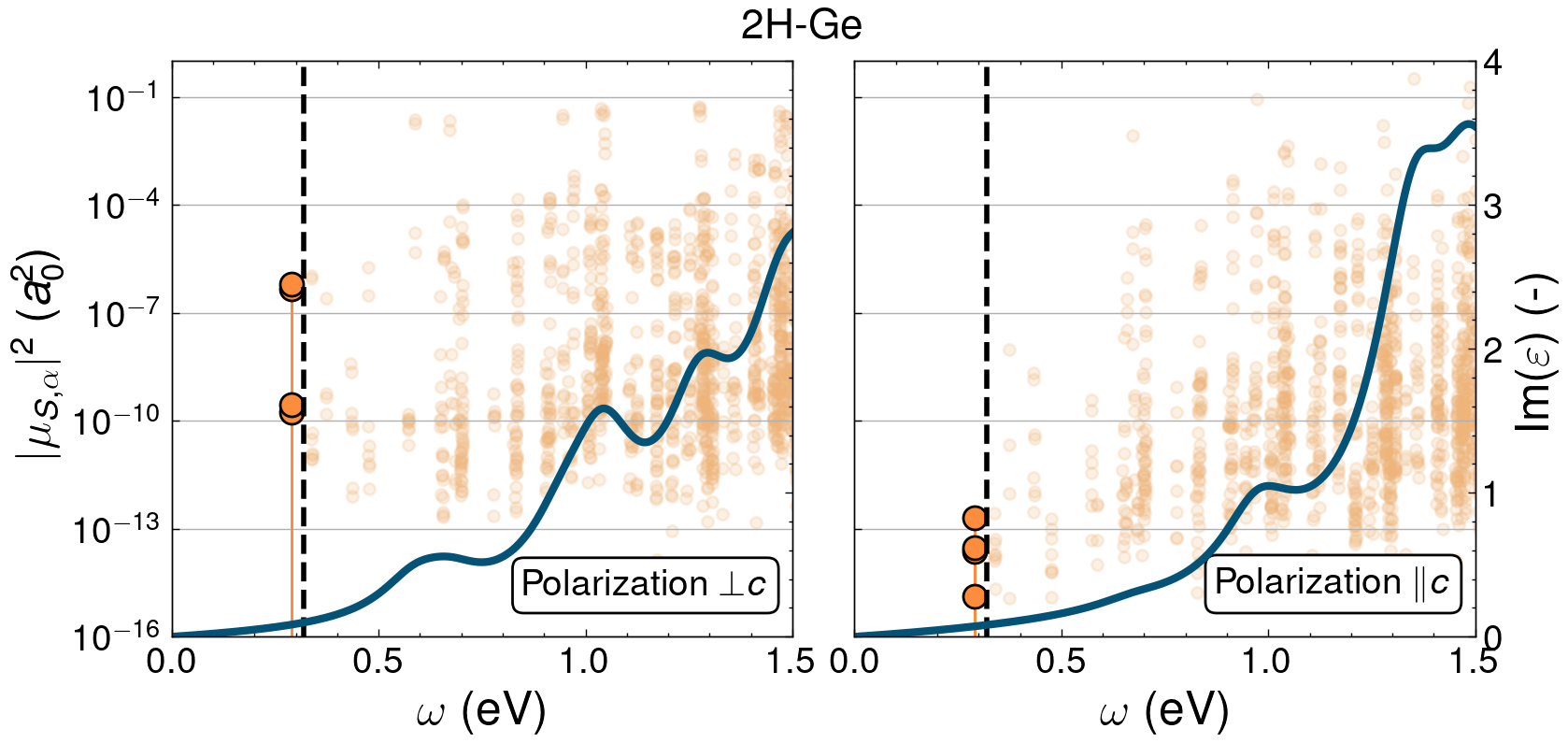}
    \caption{}
\end{subfigure}
\hfill
\begin{subfigure}[b]{0.49\linewidth}
    \includegraphics[width=\linewidth]{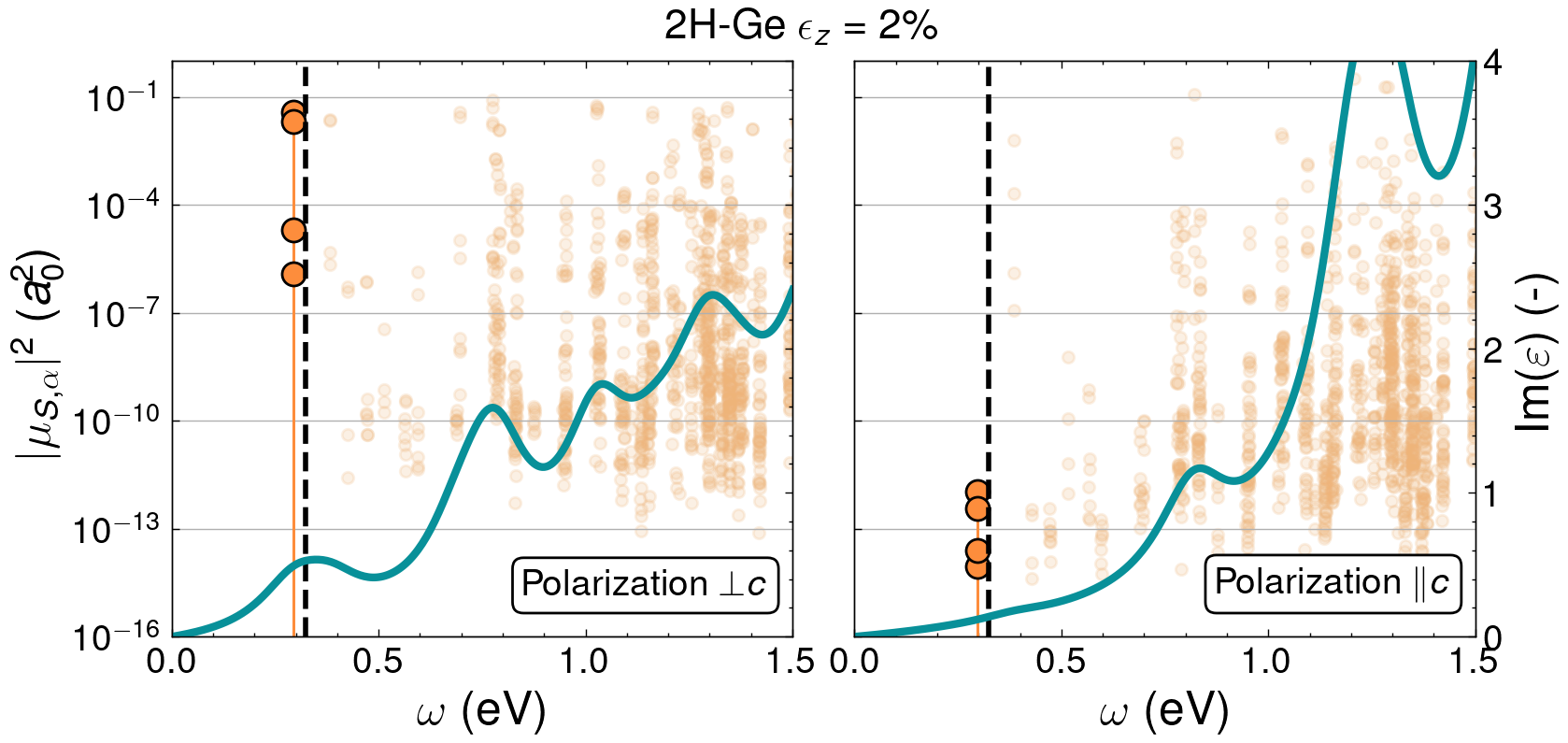}
    \caption{}
\end{subfigure}\\
\begin{subfigure}[b]{0.49\linewidth}
    \includegraphics[width=\linewidth]{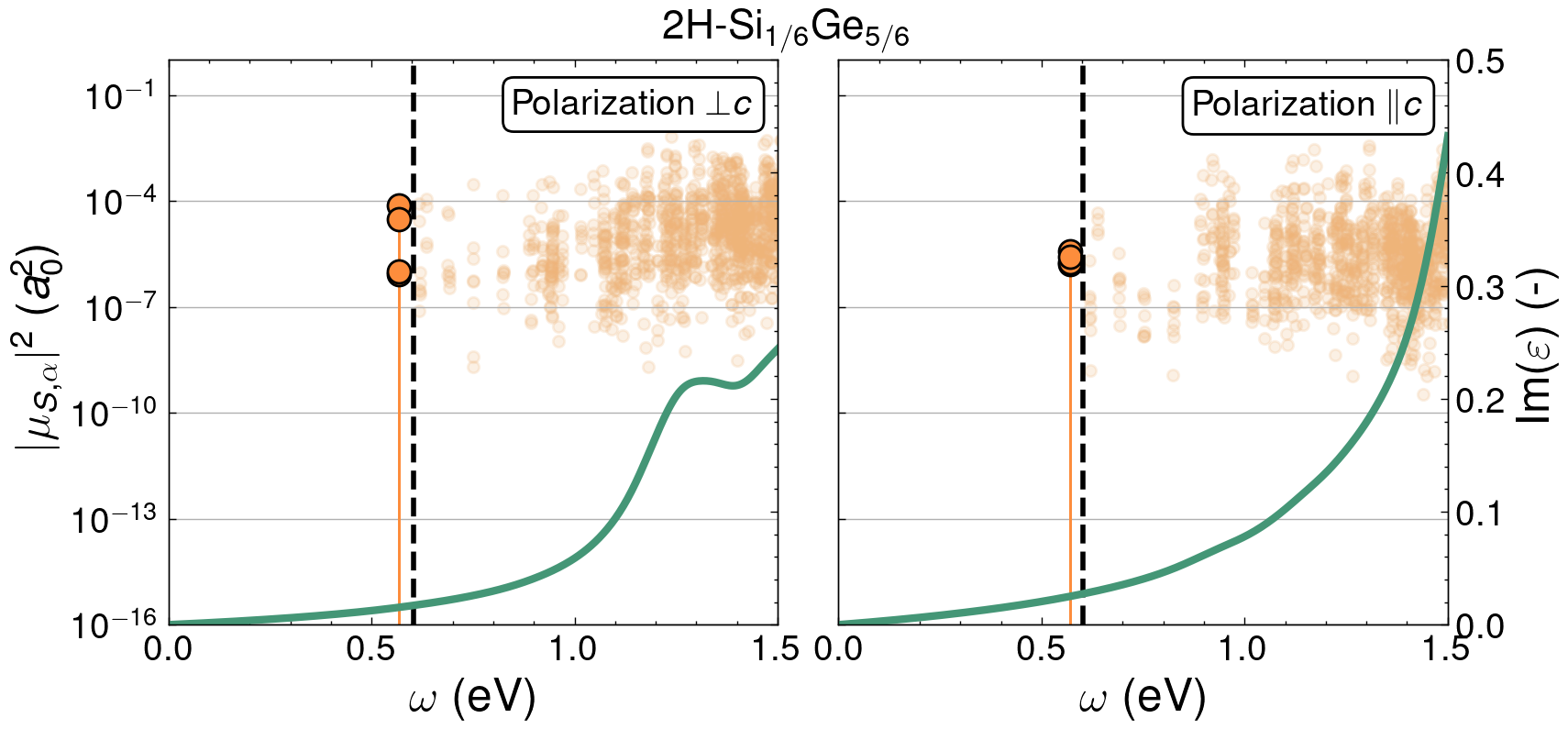}
    \caption{}
\end{subfigure}
\hfill
\begin{subfigure}[b]{0.49\linewidth}
    \includegraphics[width=\linewidth]{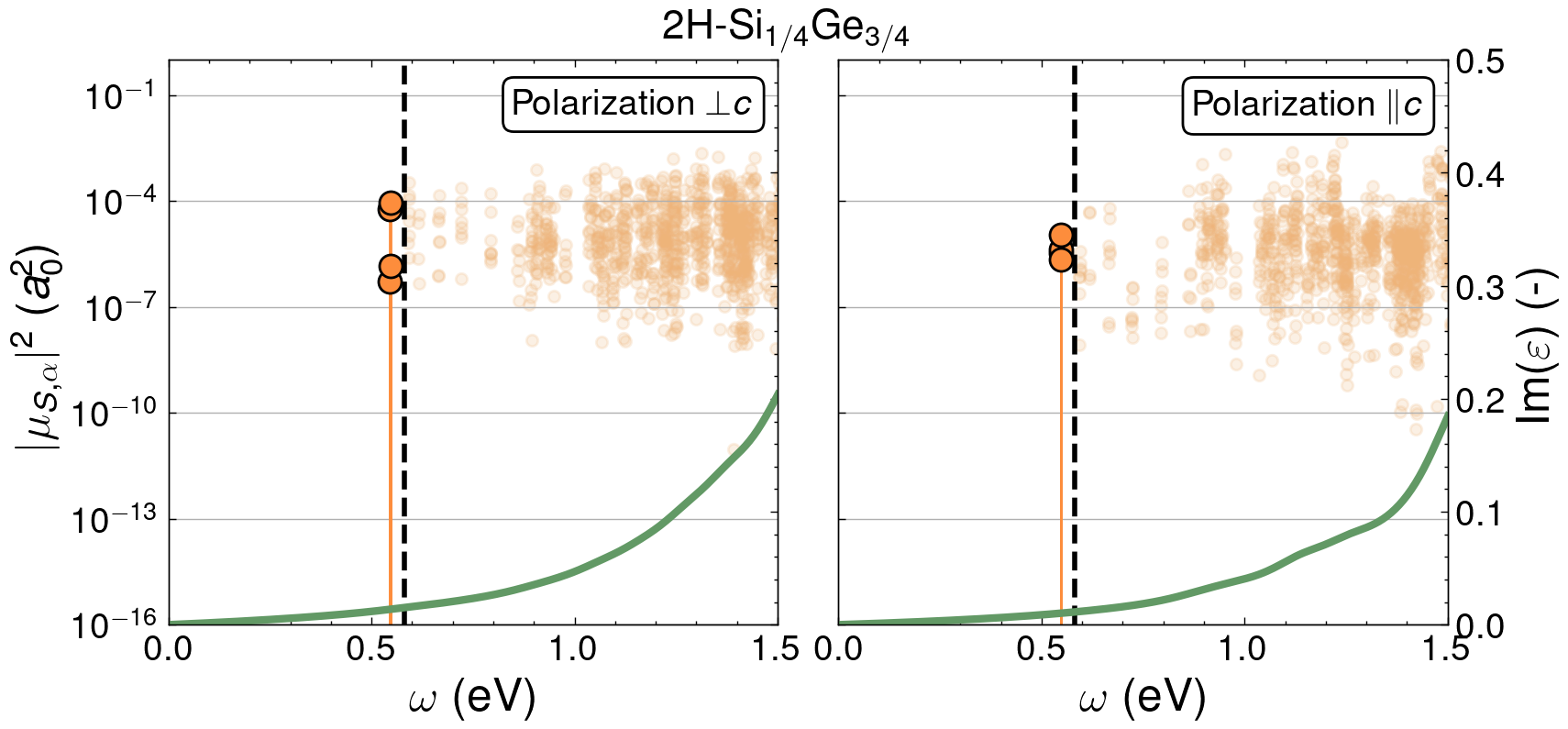}
    \caption{}
\end{subfigure}\\
\begin{subfigure}[b]{0.49\linewidth}
    \includegraphics[width=\linewidth]{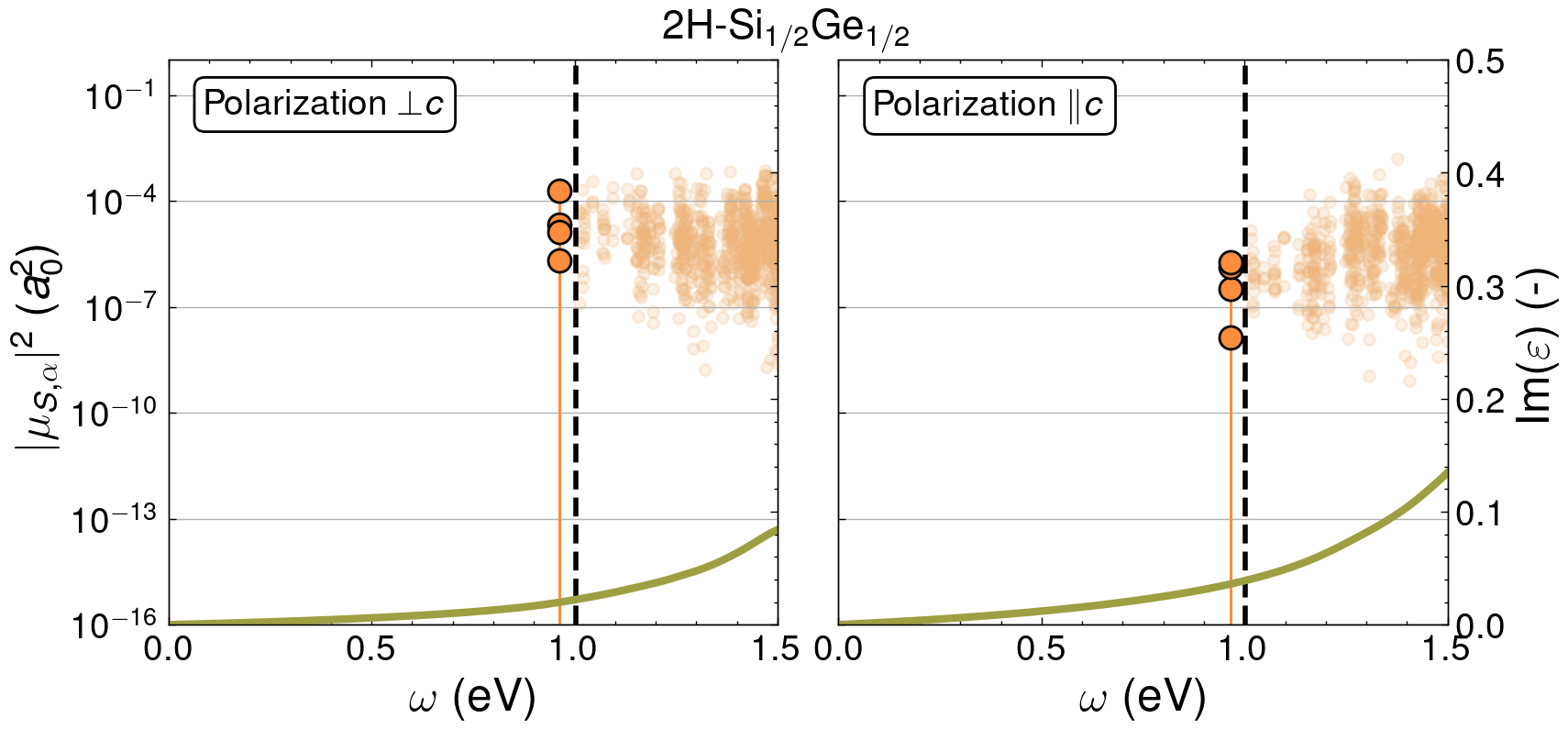}
    \caption{}
\end{subfigure}
\hfill
\begin{subfigure}[b]{0.49\linewidth}
    \includegraphics[width=\linewidth]{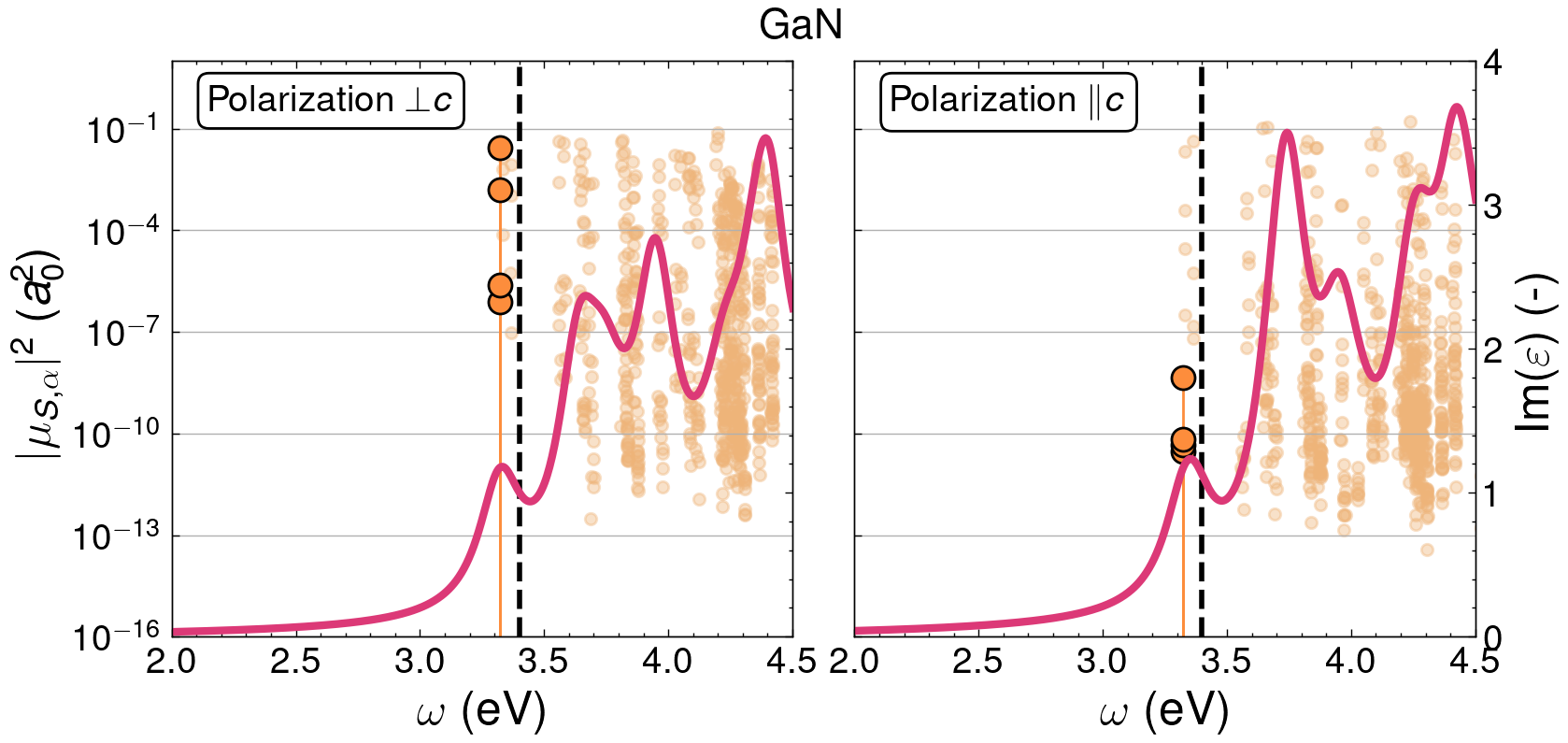}
    \caption{}
\end{subfigure}
\caption{Absorption spectra, $\mathrm{Im}(\varepsilon)$, and squared moduli of the exciton dipole moments, $|\mu_{S,\alpha}|^2$, in units of the Bohr radius squared, for 2H-Ge in panel (a), 2H-Ge under $\epsilon_z = 2\%$ uniaxial strain in (b), the 2H-Si$_{\nicefrac{1}{6}}$Ge$_{\nicefrac{5}{6}}$ alloy in (c), the 2H-Si$_{\nicefrac{1}{4}}$Ge$_{\nicefrac{3}{4}}$ alloy in (d), the 2H-Si$_{\nicefrac{1}{2}}$Ge$_{\nicefrac{1}{2}}$ alloy in (e), and GaN in (f). For each material, results are shown for in-plane ($\bot c$) and out-of-plane ($\| c$) light polarizations. The dashed vertical lines indicate the electronic bandgaps.
}
\label{fig:dipoles}
\end{figure*}

The solution of the BSE provides direct access to the optical absorption spectrum, via the imaginary part of the dielectric function, $\mathrm{Im}(\varepsilon)$, as well as to the excitonic fine structure, including excitation energies and dipole moments of the individual states, Eq.~(\ref{eq:dipoles}). Owing to the anisotropy of the hexagonal lattice, we distinguish between light polarized in the basal plane ($\bot c$) and along the $c$ axis ($\| c$). The corresponding absorption spectra and squared moduli of the exciton dipole moments $|\mu_{S,\alpha}|^2$ for pristine 2H-Ge are shown in Fig.~\ref{fig:dipoles}(a). The solid lines report $\mathrm{Im}(\varepsilon_\alpha)$ for in-plane (left panel) and out-of-plane (right panel) polarized light, while the orange dots indicate the energies and the $|\mu_{S,\alpha}|^2$, expressed in units of the Bohr radius squared ($a_0^2$), for all excitonic states obtained from the BSE. The lowest-energy excitons are highlighted for clarity.

The lowest-energy exciton state is four-fold degenerate, originating from transitions between the twofold-degenerate VBM and CBm at $\Gamma$. Its energy, $E_{1-4}$, lies about 31~meV below the electronic bandgap, as reported in Tab.~\ref{tab:oscillator}, indicating a sizable binding energy that remains relevant up to room temperature. 
The comparatively large binding energy in 2H-Ge can be rationalized within a Wannier–Mott picture, considering that the reduced electron–hole mass along the $c$ axis is  substantially larger than in 3C-Ge, and that the 2H phase exhibits slightly lower optical dielectric constants, computed to be $\varepsilon_{\bot c}=14.35,\,\varepsilon_{\|c}=14.84$, in comparison to $\varepsilon=15.8$ in 3C-Ge~\cite{Dunlap_PhysRev_1953}. These factors naturally account for the stronger exciton binding in 2H-Ge compared to the 3C phase.
This justifies the explicit inclusion of electron–hole interactions via the BSE and the analysis of intrinsically excitonic quantities such as dipole moments and radiative lifetimes.
The dipole moments of the four degenerate lowest-energy excitons for in-plane polarization range from $10^{-10}$ to $10^{-7}\,a_0^2$, whereas for out-of-plane polarization they are several orders of magnitude smaller. These states are therefore dark for light polarized $\| c$ and only weakly active for $\bot c$ polarization. For comparison, Fig.~\ref{fig:dipoles}(f) reports the corresponding $|\mu_{S,\alpha}|^2$ values for GaN, a well-known efficient light emitter: its in-plane dipole moments reach values up to $10^{-1}\,a_0^2$, more than six orders of magnitude larger than those of 2H-Ge. This confirms that radiative emission in 2H-Ge is not only strongly in-plane polarized but intrinsically very weak.
Consistently, the absorption spectrum of 2H-Ge shows no noticeable structure at the band edge, in agreement with previous calculations \cite{Tizei_NanoLett_2025}, reflecting the extremely small dipole moment of the lowest exciton. The corresponding dimensionless oscillator strength, evaluated according to Eq.~(\ref{eq:oscillator}) and therefore incorporating both the dipole moment and the exciton energy, reaches a maximum value of $6.61\cdot 10^{-9}$ for in-plane polarized light (Tab.~\ref{tab:oscillator}). Only above $\sim$0.5~eV a visible feature appears in the spectrum, followed by a steady increase beyond 1~eV, driven by strong transitions involving the VBM and the CBm+1 states.
\begin{table}[b]
\caption{%
Energies of the four degenerate lowest-energy exciton states, $E_{1-4}$ (eV), and maximum oscillator-strength values (dimensionless), computed via Eq.~(\ref{eq:oscillator}), among the degenerate states for the two light polarizations.
}
\label{tab:oscillator}
\begin{ruledtabular}
\begin{tabular}{lc|cc}
& $E_{1-4}$ & $f^\mathrm{max}_{1-4,\bot c}$ & $f^\mathrm{max}_{1-4,\| c}$\\
\colrule
2H-Ge & 0.289 & $6.61\cdot10^{-9}$ & $2.19\cdot10^{-15}$\\
2H-Ge $\epsilon_z=2\%$ & 0.295 & $3.99\cdot10^{-4}$ & $1.15\cdot10^{-14}$ \\
2H-Si$_{\nicefrac{1}{6}}$Ge$_{\nicefrac{5}{6}}$ & 0.568 & $1.59\cdot10^{-6}$ & $7.73\cdot10^{-8}$\\
2H-Si$_{\nicefrac{1}{4}}$Ge$_{\nicefrac{3}{4}}$ & 0.546 & $1.80\cdot10^{-6}$ & $8.57\cdot10^{-8}$\\
2H-Si$_{\nicefrac{1}{2}}$Ge$_{\nicefrac{1}{2}}$ & 0.962 & $7.08\cdot10^{-6}$ & $6.64\cdot10^{-8}$\\[5pt]
GaN & 3.322 & $3.41\cdot10^{-3}$ & $5.61\cdot10^{-10}$\\
\end{tabular}
\end{ruledtabular}
\end{table}

The picture changes markedly in uniaxially strained 2H-Ge, Fig.~\ref{fig:dipoles}(b). As discussed above, the band crossover activates much stronger transitions from the VBM ($\Gamma_{9v}^+$) to the new CBm ($\Gamma^-_{7c}$), which now dominate the lowest-energy excitons. The dipole moments for $\| c$ polarization are only weakly affected, whereas those for $\bot c$ polarization increase by more than five orders of magnitude relative to the unstrained material. Under $\epsilon_z = 2\%$ strain, the lowest-energy excitons reach dipole moments around $10^{-1}\,a_0^{2}$, comparable to GaN.
In contrast to the unstrained case, the absorption spectrum now exhibits a clear feature at the band edge before following the same overall trend as pristine 2H-Ge and increasing steadily at higher energies. The maximum oscillator strength accordingly reaches $3.99\cdot10^{-4}$ for $\bot c$ polarization (Tab.~\ref{tab:oscillator}). This confirms that applying about $2\%$ uniaxial strain along the $c$ axis substantially enhances the light-matter interactions of 2H-Ge through the strain-induced band crossover.

The alloy systems, shown in Figs.~\ref{fig:dipoles}(c)–(e), all exhibit a slow onset of absorption starting above $\sim\!0.5$~eV and reaching only about one tenth of the intensity observed in pristine and strained 2H-Ge around 1.5~eV (cf.~the vertical scales in panels (a)--(e)). The absorption increases more slowly with increasing Si content. Spectral features that are visible for $x=\nicefrac{1}{6}$ above 1~eV for $\bot c$ polarization become increasingly suppressed at higher Si concentrations and eventually disappear.
This trend originates from the pronounced progressive upward shift of the CBm+1 state at the $\Gamma$ point, see Figs.~\ref{fig:bands}(c)--(e), which pushes the strong optical transitions driving the absorption to significantly higher energies than in the 2H-Ge systems.
Alloying also markedly reduces the contrast between in-plane and out-of-plane dipole strengths. For both polarizations, the exciton dipole moments span the range $10^{-7}$ -- $10^{-4}\,a_0^{2}$, with the lowest-energy states reaching values of order $10^{-4}\,a_0^{2}$ for $\bot c$ and $10^{-5}\,a_0^{2}$ for $\| c$. This behavior is consistent across the different Si concentrations considered here, with comparable dipole-moment magnitudes observed throughout the explored composition range.
The symmetry breaking introduced by the random distribution of Si atoms lifts the suppression of the $\| c$ dipole component present in pristine 2H-Ge. As a result, the alloys exhibit optical responses that are nearly isotropic with respect to light polarization, with comparable dipole strengths for the $\bot c$ and $\| c$ directions. 
We note that this effect may be enhanced by the finite size of the alloy supercells considered here, where the reduced symmetry of a single representative configuration can partially overestimate polarization mixing compared to the macroscopic random-alloy limit.
The corresponding oscillator strengths of the lowest-energy states fall between the very small values found in pristine 2H-Ge and the significantly larger values characteristic of uniaxially strained 2H-Ge.

The results for our GaN comparison are shown in Fig.~\ref{fig:dipoles}(f). The binding energy of the lowest-energy, fourfold-degenerate excitons is larger than in the 2H-Ge–based systems, in agreement with previous calculations~\cite{Laskowski_PRB_2005}.
As noted earlier, GaN is an efficient light-emitting semiconductor and, correspondingly, the in-plane exciton dipole moments are of order $10^{-1}\,a_0^{2}$, whereas the out-of-plane components are about seven orders of magnitude smaller. The combination of these large $\bot c$ dipole moments with the substantially higher exciton energy $E_{1-4}=3.322$~eV yields oscillator strengths that exceed those of strained 2H-Ge by approximately one order of magnitude, reaching $3.41\cdot10^{-3}$ along the $\bot c$ direction (Tab.~\ref{tab:oscillator}).

\subsection{Exciton radiative lifetimes}
\begin{figure}[h]
    \centering
    \includegraphics[width=\linewidth]{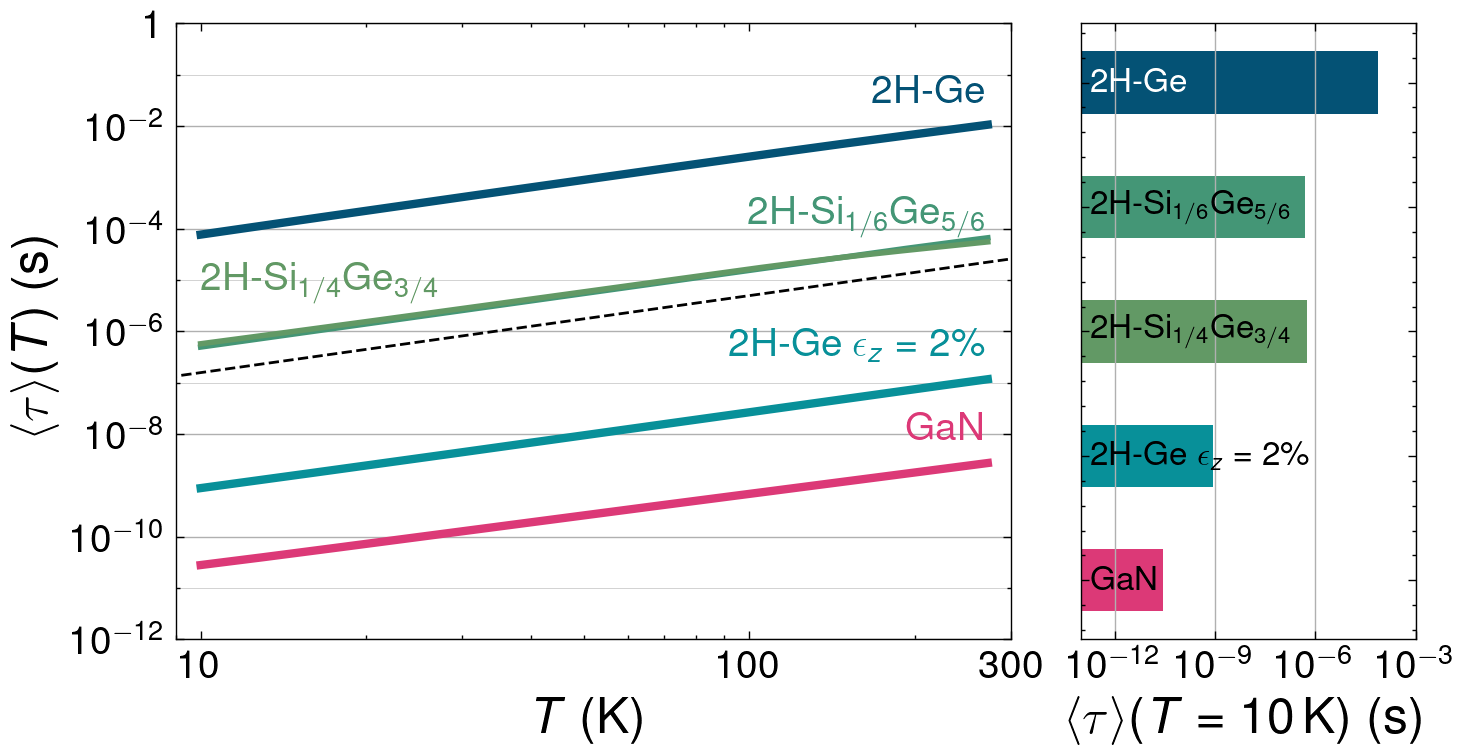}
    \caption{Temperature-averaged exciton radiative lifetimes, Eq.~(\ref{eq:tau_avg}), of the studied materials. Left panel, from top to bottom: temperature dependence of $\langle\tau\rangle(T)$ for pristine 2H-Ge, 2H-Si$_{\nicefrac{1}{6}}$Ge$_{\nicefrac{5}{6}}$ and 2H-Si$_{\nicefrac{1}{4}}$Ge$_{\nicefrac{3}{4}}$ alloys, uniaxially strained 2H-Ge and GaN. The dashed black line shows an exact $T^{3/2}$ trend. Right panel: comparison of the materials' average radiative lifetimes at $T=10$~K.}
    \label{fig:lifetimes}
\end{figure}

From the exciton dipole moments along the two polarization directions, and using Eqs.~(\ref{eq:tauS}) and (\ref{eq:tau_avg}), we compute the thermally averaged radiative lifetimes $\langle\tau\rangle(T)$ of the studied systems, shown in Fig.~\ref{fig:lifetimes}. The left panel reports the temperature dependence of $\langle\tau\rangle(T)$ for the different materials, while the right panel compares the average lifetimes evaluated at $T=10$~K. The 2H-Si$_{\nicefrac{1}{2}}$Ge$_{\nicefrac{1}{2}}$ alloy is not considered, as its electronic bandgap is indirect and the lowest-energy exciton is indirect, rendering radiative recombination momentum-forbidden.

All lifetimes reported here are computed in the intrinsic limit, i.e. in the absence of doping or free carriers, so that excitons remain stable and unscreened throughout the temperature range and no extrinsic recombination channels or many-body screening effects alter the radiative dynamics. 

Across all materials, the average lifetimes follow the $T^{3/2}$ scaling, indicated by the dashed black line in the left panel of Fig.~\ref{fig:lifetimes}, with remarkable accuracy. This behavior reflects the fact that, up to room temperature, the radiative dynamics are governed almost entirely by the fourfold-degenerate lowest-energy excitons. In all 2H-Ge--based systems, the next excitonic states lie more than 45~meV higher in energy and therefore remain essentially unpopulated over the explored temperature range. As a result, the thermal average in Eq.~(\ref{eq:tau_avg}) reduces to the contribution of these four states alone, without any progressive activation of higher excitons. The temperature dependence of $\langle\tau\rangle(T)$ therefore arises solely from the $T$ dependence of the single-state lifetime $\tau_S(T)$, i.e., from the thermal factor in Eq.~(\ref{eq:tauS}), leading to the characteristic $T^{3/2}$ scaling.

Pristine 2H-Ge exhibits the longest radiative lifetime among all materials considered, already exceeding by two orders of magnitude the microsecond scale at low temperature. This originates from the exceptionally small exciton dipole moments for both polarization directions and confirms the very low light-emission efficiency of hexagonal germanium. If the electron--hole interaction is neglected, i.e., by switching off the direct and exchange terms in the BSE kernel, while still enforcing momentum conservation within the light cone, the resulting radiative lifetime increases by roughly a factor two, as shown in the Supplemental Material, Fig.~S4.

Although the excitonic and the fully IP regimes represent fundamentally different physical pictures, a comparison remains useful. At low temperatures ($\sim 10$~K), the excitonic radiative lifetime obtained here agrees well with the IP values reported in Refs.~\cite{Rodl_PRM_2019,Fadaly_Nature_2020}, both being of order $10^{-4}\,\unit{s}$. This correspondence arises because, in this regime, only the lowest degenerate exciton is thermally occupied and its CMM remains extremely close to zero. The exciton wavefunction is largely dominated by a single interband transition at $\Gamma$, and radiative recombination effectively reduces to the same vertical transition captured in the IP picture.
As mentioned in Section~\ref{sec:methods}, at higher temperatures, however, the two descriptions diverge due to the different treatment of momentum conservation in radiative recombination that is employed: in the excitonic case, the thermal occupation of finite-momentum excitons, together with the light-cone constraint, leads to the characteristic $T^{3/2}$ increase of the radiative lifetime, whereas in the IP picture the thermal population of higher-energy vertical transitions causes a reduction of the radiative lifetime, since the light-cone constraint is not explicitly enforced.

The increase in exciton dipole moments induced by alloying with Si leads to a reduction of the radiative lifetime by roughly two orders of magnitude, bringing it down to the microsecond range at $T\!\sim\!10$~K for both the 2H-Si$_{\nicefrac{1}{6}}$Ge$_{\nicefrac{5}{6}}$ and 2H-Si$_{\nicefrac{1}{4}}$Ge$_{\nicefrac{3}{4}}$ alloys (see the right panel of Fig.~\ref{fig:lifetimes}). Correspondingly, their $\langle\tau\rangle(T)$ values overlap in the left panel of Fig.~\ref{fig:lifetimes}, indicating that no substantial difference in radiative efficiency emerges  at these alloy compositions. A much stronger reduction is achieved through the application of uniaxial strain along the $c$ axis. Under $\epsilon_z = 2\%$ strain, the average radiative lifetime decreases by about five orders of magnitude compared to pristine 2H-Ge, reaching the nanosecond scale at low temperature. This reduction directly reflects the bright in-plane transitions enabled by the strain-induced band crossover, placing strained 2H-Ge within technologically relevant emission timescales.

The radiative lifetime of strained 2H-Ge remains, however, about one order of magnitude longer than that of our GaN benchmark, which reaches $\sim\!\!30\,\unit{ps}$ at 10~K, in excellent agreement with the literature~\cite{Im_APL_1997,Jhalani_JPhysCondMat_2020}. Although the exciton dipole moments for $\bot c$ polarization are of similar magnitude in strained 2H-Ge and GaN, the shorter GaN lifetime originates from its much larger exciton energy: the energy of the degenerate lowest states in GaN is nearly an order of magnitude higher, which directly amplifies the oscillator strength (cf.\ Table~\ref{tab:oscillator}) and correspondingly reduces the radiative lifetime, see Eq.~(\ref{eq:tauS}).

\section{Summary and Conclusion}
\label{sec:conclusion}
In this work we present a comprehensive excitonic analysis of pristine 2H-Ge and of its alloyed and strained variants, using wurtzite GaN as reference. By combining \textit{ab initio} DFT$+J$ electronic structures with BSE calculations, we quantify the exciton binding energies, dipole moments, and intrinsic radiative lifetimes of these systems.

Across all 2H-Ge–based materials studied, the sizeable binding energies demonstrate that excitonic effects remain relevant well above cryogenic temperatures and must be explicitly included to capture the optical response. The fourfold-degenerate excitons at the band edge dominate the radiative dynamics up to room temperature, leading to the characteristic $T^{3/2}$ dependence of the average radiative lifetime.

Our results show that pristine 2H-Ge possesses exceptionally small exciton dipole moments for both in-plane and out-of-plane light polarization, resulting in very long radiative lifetimes exceeding $10^{-4}\,\unit{s}$ at low temperature. This confirms the intrinsically weak optical activity of hexagonal Ge and is consistent with the absence of strong spectral features near the band edge reported experimentally.

Alloying Ge with Si moderately increases the dipole strength and reduces the radiative lifetime, primarily due to the lifted constraints on optical transitions. A much more pronounced enhancement arises under uniaxial strain along the $c$ axis, where the strain-induced band crossover activates strong in-plane transitions, yielding dipole moments and radiative efficiencies comparable to those of GaN. Under $\epsilon_z = 2\%$ strain, the radiative lifetime reaches the nanosecond regime, with a five-orders of magnitude improvement over pristine 2H-Ge.

Taken together, our results provide an additional investigation on the intrinsic optical response of 2H-Ge, reinforcing earlier theoretical indications of a mismatch between its band-edge properties and the strong PL reported in some experiments. The very weak exciton dipoles and long intrinsic lifetimes obtained here support the view that the observed PL is likely mediated by extrinsic channels, such as defects, morphology or local strain fields, rather than by the radiative recombination of a bright exciton. Conversely, our results identify strain engineering as a highly effective route to enhance intrinsic emission in hexagonal Ge, far more efficient than alloying.

Overall, this work fills a gap in the current understanding of hexagonal Ge by providing a microscopic excitonic perspective on its radiative properties and offering additional quantitative evidence that clarifies the intrinsic limits of light emission in this polytype.
\bigskip

\begin{acknowledgments}
MRF and MP thank Ivan Marri and Sudha Priyanga Ganesapandian for helpful discussions. MRF and MP acknowledge the ``Italian Research Center on High Performance Computing, Big Data and Quantum Computing'' (ICSC) funded by the European Union-NextGenerationEU and established under the National Recovery and Resilience Plan (PNRR), as well as high-performance computing resources provided by CINECA through the ISCRA initiative. MA acknowledges the ANR AMPHORE project (ANR-21-CE09-0007) and the ANR TULIP (ANR-24-CE09-5076). MP thanks INFN for the TIME2QUEST funding project.
\end{acknowledgments}

\bibliography{refs}

\end{document}